\documentclass[apj]{emulateapj}
\usepackage{color}
\usepackage{ulem}

\slugcomment{accepted version (9th November 2014)}
\shorttitle{}
\shortauthors{Sano et al.}

\begin{document}
\title{A DETAILED STUDY OF NON-THERMAL X-RAY PROPERTIES AND INTERSTELLAR GAS TOWARD THE $\gamma$-RAY SUPERNOVA REMNANT RX J1713.7$-$3946}

\author{H. Sano\altaffilmark{1}, T. Fukuda\altaffilmark{1}, S. Yoshiike\altaffilmark{1}, J. Sato\altaffilmark{1}, H. Horachi\altaffilmark{1}, T. Kuwahara\altaffilmark{1}, K. Torii\altaffilmark{1}, T. Hayakawa\altaffilmark{1}, T. Tanaka\altaffilmark{2}, H. Matsumoto\altaffilmark{1}, T. Inoue\altaffilmark{3}, R. Yamazaki\altaffilmark{3}, S. Inutsuka\altaffilmark{1}, A. Kawamura\altaffilmark{4}, H. Yamamoto\altaffilmark{1}, T. Okuda\altaffilmark{4}, K. Tachihara\altaffilmark{1}, N. Mizuno\altaffilmark{4}, T. Onishi\altaffilmark{5}, A. Mizuno\altaffilmark{6}, F. Acero\altaffilmark{7}, and Y. Fukui\altaffilmark{1}}

\affil{$^1$Department of Physics, Nagoya University, Furo-cho, Chikusa-ku, Nagoya 464-8601, Japan; sano@a.phys.nagoya-u.ac.jp}
\affil{$^2$Department of Physics, Kyoto University, Kitashirakawa-oiwake-cho, Sakyo-ku, Kyoto 606-8502, Japan}
\affil{$^3$Department of Physics and Mathematics, Aoyama Gakuin University, Fuchinobe, Chuou-ku, Sagamihara 252-5258, Japan}
\affil{$^4$National Astronomical Observatory of Japan, Mitaka 181-8588, Japan}
\affil{$^5$Department of Astrophysics, Graduate School of Science, Osaka Prefecture University, 1-1 Gakuen-cho, Naka-ku, Sakai 599-8531, Japan}
\affil{$^6$Solar-Terrestrial Environment Laboratory, Nagoya University, Furo-cho, Chikusa-ku, Nagoya 464-8601, Japan}
\affil{$^7$LUPM, CNRS/Universit$\mathrm{\acute{e}}$ Montpellier 2, France}

\begin{abstract}
We have carried out a spectral analysis of the $Suzaku$ X-ray data in the 0.4--12 keV range toward the shell-type very-high-energy $\gamma$-ray supernova remnant RX J1713.7$-$3946. The aims of this analysis are to estimate detailed X-rays spectral properties at a high angular resolution up to 2 arcmin, and to compare them with the interstellar gas. The X-ray spectrum is non-thermal and used to calculate absorbing column density, photon index, and absorption-corrected X-ray flux. The photon index varies significantly from 2.1 to 2.9. It is shown that the X-ray intensity is well correlated with the photon index, especially in the west region, with a correlation coefficient of 0.81. The X-ray intensity tends to increase with the averaged interstellar gas density while the dispersion is relatively large. The hardest spectra having the photon index less than 2.4 are found outside of the central 10 arcmin of the SNR, from the north to the southeast ($\sim$430 arcmin$^2$) and from the southwest to the northwest ($\sim$150 arcmin$^2$). The former region shows low interstellar gas density, while the latter high interstellar gas density. We present discussion for possible scenarios which explain the distribution of the photon index and its relationship with the interstellar gas.
\end{abstract}


\section{Introduction}
Supernova remnants (SNRs) in the Galaxy are believed to be most likely accelerators of cosmic-rays (CRs) below ``Knee\rq{}\rq{}, in an energy range less than $3\times$10$^{15}$ eV. The highest energy $\gamma$-rays in the Galaxy are emitted from three young and very-high-energy (VHE; $E$ $>$ 100 GeV) $\gamma$-ray SNRs, RX J1713.7$-$3946 \citep{aharonian2004,aharonian2006b,aharonian2007a}, RX J0852.0$-$4622 \citep{aharonian2005,aharonian2007b}, and HESS J1731$-$347 \citep{hess2011}. In these SNRs, it is suggested that CR protons or electrons are accelerated efficiently up to 10--800 TeV or 1--40 TeV, respectively \citep[e.g.,][]{zirakashvili2010}. All the three show pure synchrotron X-rays with no thermal features, and acceleration of TeV CR electrons is well established. The origin of the $\gamma$-rays in these SNRs, either hadronic or leptonic, has been under debate; the GeV $\gamma$-ray observations show hard $\gamma$-ray spectrum similar to that of the leptonic $\gamma$-rays \citep{abdo2011}, whereas a recent comparative study between $\gamma$-rays and the interstellar protons is consistent with the hadronic origin of $\gamma$-rays \citep{fukui2012,fukui2013}. It is notable that the numerical simulations on the shock-cloud interaction \citep{inoue2012} lend support for the hadronic $\gamma$-rays, since the low energy CR protons producing the GeV $\gamma$-rays cannot penetrate into the densest part of the cloud cores, leading to the hard spectrum just presented by \cite{abdo2011}. A similar conclusion is drawn by \cite{zirakashvili2010} and \cite{gabici2014}. Even though the Leptonic scenarios are still debated from a different perspective \citep[e.g.,][]{ellison2012}.

The young SNR RX J1713.7$-$3946 (also known as G347.3$-$0.5) is one of the best targets for testing the action of the CR electrons, because the SNR emits strong synchrotron X-rays and VHE $\gamma$-rays \citep{pfeffermann1996,enomoto2002,aharonian2004,aharonian2006a,aharonian2006b,aharonian2007a}. RX J1713.7$-$3946 has a large apparent diameter of $\sim$1 degree at a small distance of $\sim$1 kpc \citep{fukui2003}. Subsequently, the X-ray properties of RX J1713.7$-$3946 were studied by using $XMM$-$Newton$, $Suzaku$, and $Chandra$. $Chandra$ observations revealed filamentary structures of X-rays  and short time variability ($\sim$1 yr) of one of them \citep{uchiyama2003,uchiyama2007}, which is interpreted as due to rapid particle acceleration of CR electrons and cooling in $\sim$1 yr scale with mG magnetic field. If this is correct, the CR electrons should be accelerated efficiently in the ``Bohm diffusion regime\rq{}\rq{} \citep{uchiyama2007}. Additionally, when the cosmic-rays are accelerated, the magnetic fields amplified by cosmic-ray streaming instability \citep{lucek2000}. Moreover, different processes of amplifying magnetic fields may be working in during shock-cloud interaction \citep[e.g.,][]{giacalone2007,inoue2009,inoue2012}. The amplified magnetic field may intensify the non-thermal X-rays, whereas the synchrotron loss due to the amplified field may decrease the number of CR electrons via synchrotron loss \cite[e.g.,][]{kishishita2013}. In order to test such actions, we need a detailed comparison between the electron spectrum and the interstellar matte. \cite{cassamchenai2004} presented spatial distribution of absorbing column density $N_{\rm{H}}$(X-ray) and photon index $\Gamma$ obtained by $XMM$-$Newton$ at angular resolutions smoothed to 2--30 arcmin. The $N_{\rm{H}}$(X-ray) distribution shows significant variations over the whole SNR (0.4 $\times$ 10$^{22}$ cm$^{-2}$ $\le$ $N_{\rm{H}}$(X-ray) $\le$ 1.1 $\times$ 10$^{22}$ cm$^{-2}$). The authors carried out an analysis of X-ray absorption and compared the X-rays with CO, H{\sc i}, and visual extinction and showed that the interstellar medium (ISM) is correlated with the $N_{\rm{H}}$(X-ray). Additionally, the value of photon index $\Gamma$ has also shown the strong variation (1.8 $\leq \Gamma \leq$ 2.6). This trend was confirmed by a more detailed analysis of $XMM$-$Newton$ data \citep{hiraga2005,acero2009}. \cite{takahashi2008} and \cite{tanaka2008} detected the hard synchrotron X-rays up to 40 keV from the SNR by using $Suzaku$. The spectra are fitted by interstellar absorbed power-law model with the exponential rolloff, suggesting CR electrons close to the Bohm diffusion limit \citep{tanaka2008}.

\cite{sano2010} presented multi-transition CO data and comparing with the $Suzaku$ X-rays toward the northwest of RX J1713.7$-$3946. These authors found that the X-rays and molecular clumps are well-correlated at 1 pc scale, but anti-correlated at 0.1 pc scale. The latter shows a trend as rim-brightened X-rays around the dense molecular clumps. Subsequently, \cite{sano2013} compared 18 CO clumps and a H{\sc i} clump with their surrounding X-rays and concluded that the trend is commonly found over the whole SNR. Additionally, the authors found the strong correlation between the X-ray intensity and the molecular mass interacting with the SNR. These results indicate that the synchrotron X-rays are closely related to the surrounding CO gas. To explore this correlation, we should estimate the physical parameters such as the absorbing column density $N_{\rm{H}}$(X-ray), photon index $\Gamma$ and X-ray flux, in angular resolution high enough to compare with the ISM distributions. Such quantitative comparisons with the ISM mark an important step toward understanding the spectral variation of X-rays and efficient cosmic-ray acceleration.

Our aim is to compare the spatial distribution of X-ray properties with that of the ISM clumps. For this aim, we carried out the spectral extraction and model fitting in the regions gridded at 2--8 arcmin scales over the entire SNR. In the present paper, we show a detailed comparison of the spatial distribution among $N_{\rm{H}}$(X-ray), $\Gamma$, X-ray flux and ISM both molecular and atomic gas at angular resolutions raging from 2 arcmin to 8 arcmin depending on the photon statistics. Section \ref{section:observations} gives observations and data reduction of $Suzaku$ X-rays, NANTEN CO and ATCA $\&$ Parkes H{\sc i}. Section \ref{section:results} consists of three subsections: Section \ref{typical} presents the typical spectra of X-rays; Section \ref{characterization} spatial and spectral characterization of the X-rays; and \ref{comparisonISM} a detailed comparison with the ISM. Section \ref{section:discussion} consists of three subsection: in Section \ref{absorbingcol} discusses the spatial variation of absorbing column density is discussed; Section \ref{relationship} the relationship between X-ray flux and ISM; and Section \ref{efficient} the efficient acceleration of cosmic-ray electrons. The conclusions are given in Section \ref{section:Conclusion}.

\section{Observations and Data Reductions}
\label{section:observations}
\subsection{X-rays}
\subsubsection{Details of the datasets}
We analyzed the X-ray dataset archive obtained by $Suzaku$ (Data Archives and Transmission System; DARTS at ISAS/JAXA). This dataset consists of 17 pointings taken at 2005 September (SWG; 3 pointings), 2006 September and October (AO1; 10 pointings), 2010 February (AO4: 4 pointings) and we mainly used 15 pointing of ON sources. These data were already analyzed and published elsewhere \citep{takahashi2008,tanaka2008,sano2013}. We used Software HEASoft version 6.11 with pipeline processing version 2.0 or 2.4 and with standard event selection criteria (cleaned event files).

\begin{figure}
\begin{center}
\includegraphics[width=86mm,clip]{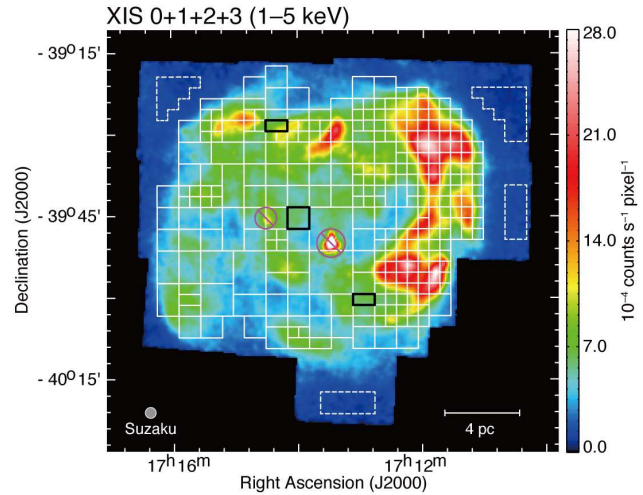}
\caption{$Suzaku$ XIS mosaic image of RX J1713.7$-$3946 in the energy band 1--5 keV \citep{sano2013}. The color scale indicates count rate on a square root scale and is unit of 10$^{-4}$ counts s$^{-1}$ pixel$^{-1}$. The solid boxes are those used for spectral analysis. We also show the spectra enclosed by the black solid lines (see also Figure \ref{fig2}). The dashed boxes correspond to the regions used to extract the background spectrum. The two point sources, 1WGA J1714.4$-$3945 and 1WGA J1713.4$-$3949, are circled with magenta and those regions were removed from spectral analysis.}
\label{fig1}
\end{center}
\end{figure}%

\subsubsection{Imaging}
The X-ray images are those used in Figure \ref{fig1}a of \cite{sano2013}. Figure \ref{fig1} shows XIS mosaic image (1--5 keV) in RX J1713.7$-$3946 \citep{sano2013}. Color scheme is in a square-root scale in 10$^{-4}$ counts s$^{-1}$ pixel$^{-1}$ (pixel size is $\sim$16.7$\arcsec$) smoothed with a Gaussian kernel with FWHM $\sim$45$\arcsec$. This image is subtracted for non X-ray background (NXB) and corrected for the vignetting effect by XRT (see more details in \citeauthor{sano2013} \citeyear{sano2013} Section 2.3).

\subsubsection{Spectroscopy}\label{section:spec}
\begin{figure}
\begin{center}
\includegraphics[width=86mm,clip]{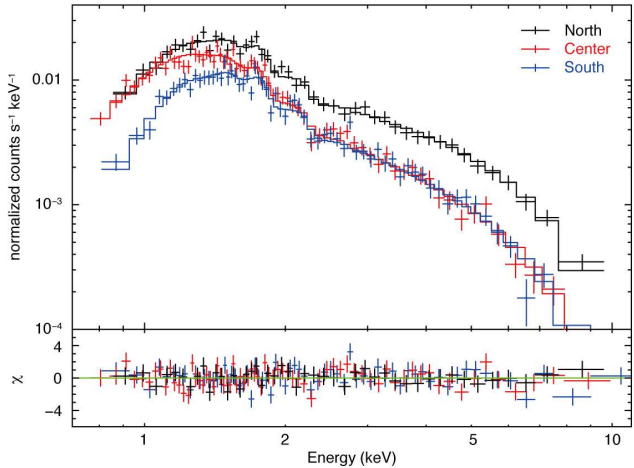}
\caption{Typical X-ray spectra of north (black), center (red) and south (blue) regions (for the regions definition see Figure \ref{fig1}). The solid lines represent the best-fit absorbed power-law model. The lower part shows the residuals from the best-fit models. The north and center spectra can be fitted by the same absorbing column density ($N_{\mathrm{H}}$(X-ray)$\sim$0.5 $\times$ 10$^{22}$ cm$^{-2}$) but with different photon index (north: $\Gamma$=2.1$\pm$0.1, center: $\Gamma$=2.7$\pm$0.1). The center and south spectra are the same photon index ($\Gamma \sim$2.7) but with different absorbing column densities (center: $N_{\mathrm{H}}$(X-ray) = 0.45$^{+0.06}_{-0.05}$ $\times$ 10$^{22}$ cm$^{-2}$, south: $N_{\mathrm{H}}$(X-ray) = 0.89$^{+0.09}_{-0.08}$ $\times$ 10$^{22}$ cm$^{-2}$).}
\label{fig2}
\end{center}
\end{figure}

The X-ray spectra of RX J1713.7$-$3946 is represented by the absorbed power-law model \citep[e.g.,][]{koyama1997}. This model is for synchrotron X-rays with photoelectric absorption and the model fitting gives three parameters, absorbing column density $N_{\rm{H}}$(X-ray), and the photon index $\Gamma$, in addition to the absorption-corrected X-ray flux. We show typical X-ray spectra in Figure \ref{fig2}. We shall explain the method to derive the three parameters above. First, the SNR was divide into $\sim$600 regions of 2$\arcmin \times 2 \arcmin$ grids, where the number of X-ray counts is more than 160 counts per grid in the energy band 1--5 keV. For the $\sim$600 regions, only the data taken with FI CCD (XIS 0, 2, and 3) were used to derive X-ray spectra. In order to compare with the data of the interstellar medium, CO and H{\sc i}, the angular resolution was set to the highest value ($Suzaku$ HPD $\sim$2$\arcmin$). In this case, the signal leaking into/out of neighboring grid cells is less than 50 $\%$, because we generated ARFs (Ancillary Response Files) with 4 arcmin grid images for each region. This signal leaking is not critical in obtaining the large-scale trend of the photon index, absorbing column density, and fluxes. In addition, the background spectrum was estimated in the four regions shown in Figure \ref{fig1} and the western background spectrum  ($\alpha_{\mathrm{J2000}} = $17$^{\mathrm{h}}$10$^{\mathrm{m}}$31$^{\mathrm{s}}$, $\delta_{\mathrm{J2000}}$ = $-$39$^{\circ}$44$\arcmin$22.7$\arcsec$) was used for every region. The two prominent point sources (1WGA J1714.4$-$3945 and 1WGA J1713.4$-$3949) in the field are excluded in the analysis. Subsequently, the spectra of each region taken with XIS 0, 2, and 3 are summed up and was fit by the absorbed power-law model to generate RMF (Redistribution Matrix File) by xisrmfgen and ARF by xissimarfgen \citep{ishisaki2007}.

Next, in order to reduce the statistical relative errors, some neighboring regions on source were summed up and the spectra were again fit by the absorbed power-law model. Specifically, we combined two or more 2 arcmin grids so that the relative error of $N_{\rm{H}}$(X-ray) becomes 30$\%$ or less. The combined grids are selected in the neighboring regions, which have roughly the same values in $N_{\rm{H}}$(X-ray).

Figure \ref{fig1} shows the final division with 305 regions as indicated by solid boxes. About one third of the whole SNR is 2$\arcmin$ grid (4 arcmin$^2$) and 80$\%$ of that is better than 4$\arcmin$ grid (16 arcmin$^2$). These values are much better than the previous studies \citep[e.g.,][70$\%$ of that is worse than 6$\arcmin$ grid (36 arcmin$^2$)]{cassamchenai2004}. Each spectrum is binned to include at least 100 counts per individual energy bin. After the binning the energy ranges a below 0.4 keV and above 12 keV bin are excluded in the fitting. The best fit parameters in the fitting were used to derive absorbing column density $N_{\rm{H}}$(X-ray), photon index $\Gamma$, and absorption-corrected flux in 3--10 keV, $F_{3-10\mathrm{keV}}$, as shown in Figure \ref{fig3}. If the background spectrum is changed to that for other three regions, the values of $N_{\rm{H}}$(X-ray) only changed systematically with 10--20$\%$ and it has no effect on the global trend in Figure \ref{fig3}.

\subsection{CO and H{\sc i}}\label{section:co_and_hi}
$^{12}$CO($J$=1--0) and H{\sc i} datasets are NANTEN Galactic Plane Survey (NGPS) and Southern Galactic Plane Survey (SGPS) taken with NANTEN, and ATCA $\&$ Parkes telescopes \citep{moriguchi2005, mccluregriffiths2005}, respectively. The angular resolutions are HPBW $\sim$2$\farcm$6 for CO and $\sim$2$\farcm$2 for H{\sc i}, similar to $Suzaku$ XIS HPD $\sim$2$\arcmin$. The velocity resolutions and typical rms noise fluctuations are (CO) 0.65 km s$^{-1}$ and 0.3 K ch$^{-1}$ and (H{\sc i}) 0.82 km s$^{-1}$ and 1.9 K ch$^{-1}$, respectively. 

The velocity integrated intensities of CO and H{\sc i}, $W$(CO) and $W$(H{\sc i}), are converted into molecular column density $N$(H$_2$) and atomic column density $N_{\rm{H}}$(H{\sc i}) and the total proton column density is obtained as $N_{\rm{H}}$(H$_2$+H{\sc i}) = 2$\times$$N$(H$_2$) + $N_{\rm{H}}$(H{\sc i}). A relationship $N$(H$_2$) (cm$^{-2}$) = $X_{\mathrm{CO}}$ (cm$^{-2}$ (K km s$^{-1}$)$^{-1}$)$\times$$W$(CO) (K km s$^{-1}$) is used where $X_{\mathrm{CO}}$ = 2.0$\times$10$^{20}$ (cm$^{-2}$ (K km s$^{-1}$)$^{-1}$) \citep{bertsch1993}. The H{\sc i} line is generally assumed to be optically thin and a relationship $N_{\rm{H}}$(H{\sc i}) = 1.823$\times$10$^{18}$ $\times$ $W$(H{\sc i}) is used \citep{dickey1990}. The regions where the H{\sc i} is optically thick and hence need to apply the correction for self-absorption. Details of the method is given by \citep{fukui2012} in Section 3.3.3.

\section{Results}\label{section:results}
\subsection{Typical X-ray spectra}\label{typical}
Figure \ref{fig2} shows $Suzaku$ XIS 0+2+3 spectra in the three typical regions, the north ($\alpha_{\mathrm{J2000}}$, $\delta_{\mathrm{J2000}}$) = (17$^{\mathrm{h}}$14$^{\mathrm{m}}$19.95$^{\mathrm{s}}$, $-$39$^{\circ}$28$\arcmin$31.11$\arcsec$), the center ($\alpha_{\mathrm{J2000}}$, $\delta_{\mathrm{J2000}}$) = (17$^{\mathrm{h}}$13$^{\mathrm{m}}$59.33$^{\mathrm{s}}$, $-$39$^{\circ}$45$\arcmin$31.49$\arcsec$), and the south ($\alpha_{\mathrm{J2000}}$, $\delta_{\mathrm{J2000}}$) = (17$^{\mathrm{h}}$12$^{\mathrm{m}}$56.75$^{\mathrm{s}}$, $-$40$^{\circ}$00$\arcmin$31.28$\arcsec$), and the results of the model fitting. The lower parts show residuals in the fitting indicating that the fitting is reasonably good. The absorbing column density $N_{\rm{H}}$(X-ray) is $\sim$0.5$\times$10$^{22}$ cm$^{-2}$ both in the north and the center, but the photon index is different as $\Gamma$ = 2.1$\pm$0.1 and $\Gamma$ = 2.7$\pm$0.1 for the north and the center, respectively. On the other hand, the south has large absorbing column density $N_{\rm{H}}$(X-ray) = 0.89$^{+0.09}_{-0.08} \times $10$^{22}$ cm$^{-2}$ and photon index $\Gamma$ $\sim$2.7. These differences are seen as significantly different spectral shapes in Figure \ref{fig2}.

\begin{figure*}
\begin{center}
\includegraphics[width=180mm,clip]{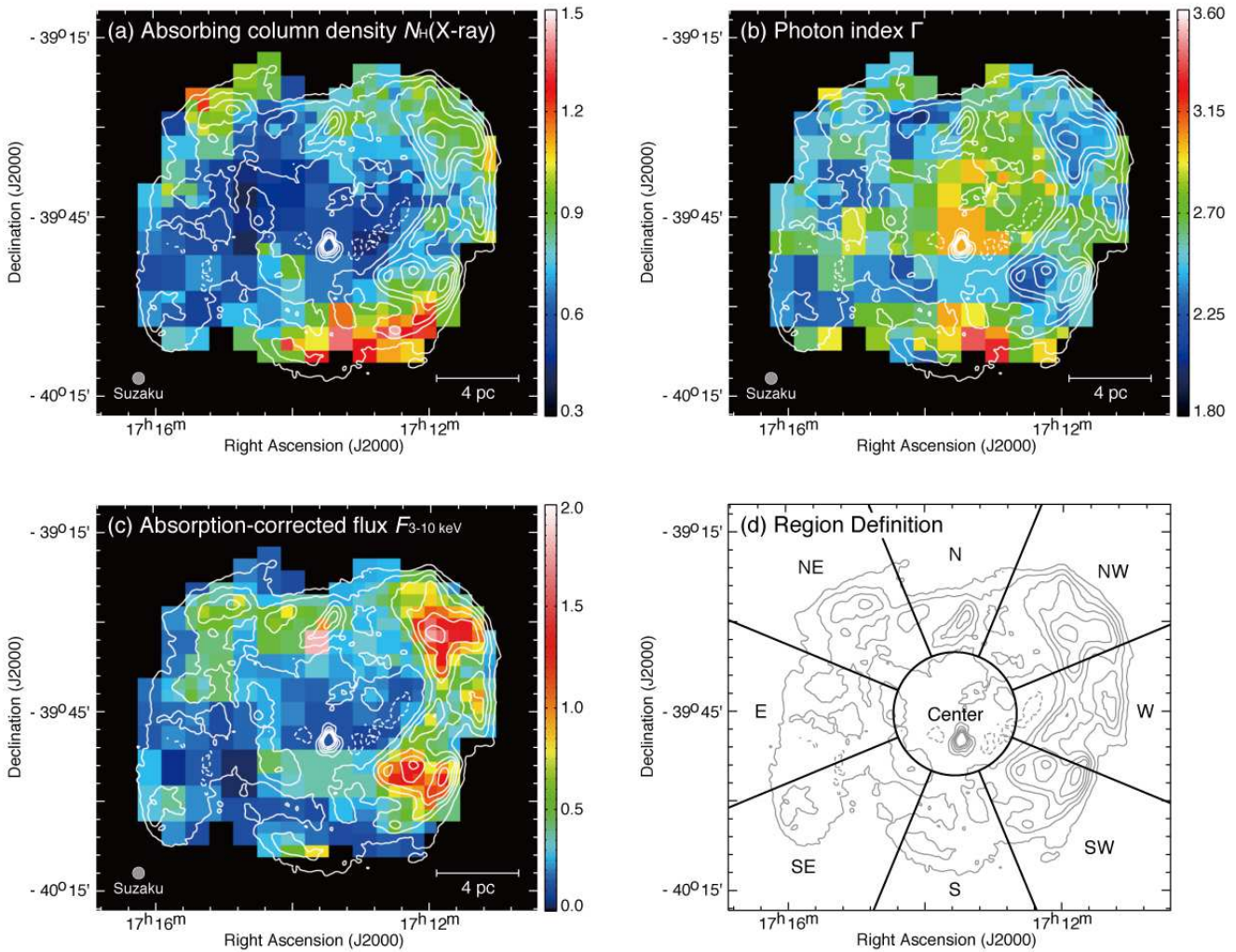}
\caption{Maps of the best-fit parameters, (a) absorbing column density $N_{\mathrm{H}}$(X-ray), (b) photon index $\Gamma$, and (c) absorption-corrected X-ray flux $F_{3-10 \mathrm{keV}}$, for absorbed power-law model. All maps overlaid with smoothed contours of the $Suzaku$ XIS mosaic images. The contour levels are from at 3.9 $\times$ 10$^{-4}$ counts s$^{-1}$ pixel$^{-1}$ and are square-root-spaced up to 23.9 $\times$ 10$^{-4}$ counts s$^{-1}$ pixel$^{-1}$ with a pixel size of $\sim$16$\farcs$7. The color schemes in (a--b) and (c) are liner and square-root-scale, respectively. The units are 10$^{22}$ cm$^{-2}$ for $N_{\mathrm{H}}$(X-ray) and $10^{-12}$ erg cm$^{-2}$ s$^{-1}$ for $F_{3-10 \mathrm{keV}}$. The definition of region names is also shown in (d). The circle centered at ($l$, $b$) = (347$\fdg$3, $-0\fdg5$) or ($\alpha_{\mathrm{J2000}}$, $\delta_{\mathrm{J2000}}$) = ($17^{\mathrm{h}}$ $13^{\mathrm{m}}$ $34^{\mathrm{s}}$, $-39^{\circ}$ 48$\arcmin$ $17\arcsec$), has 3 pc in radius.}
\label{fig3}
\end{center}
\end{figure*}%

\subsection{Spatial and spectral characterization of the X-rays}\label{characterization}

\subsubsection{Absorbing column density}
Figure \ref{fig3}a shows absorbing column density $N_{\rm{H}}$(X-ray) in the individual regions, where the $Suzaku$ XIS 1--5 keV intensity is overlayed as contours. We aimed at achieving that the relative error is 30$\%$ at maximum at the 90$\%$ confidence level by tuning the pixel size. Consequently, the average relative error and its maximum value are confined to be 14$\%$ and 30$\%$, respectively, at the 90$\%$ confidence level. This accuracy is high enough to assess the spatial variation of the absorbing column density and photon index as discussed later. We also defined the region name in Figure \ref{fig3}d.

\begin{figure*}
\begin{center}
\includegraphics[width=180mm,clip]{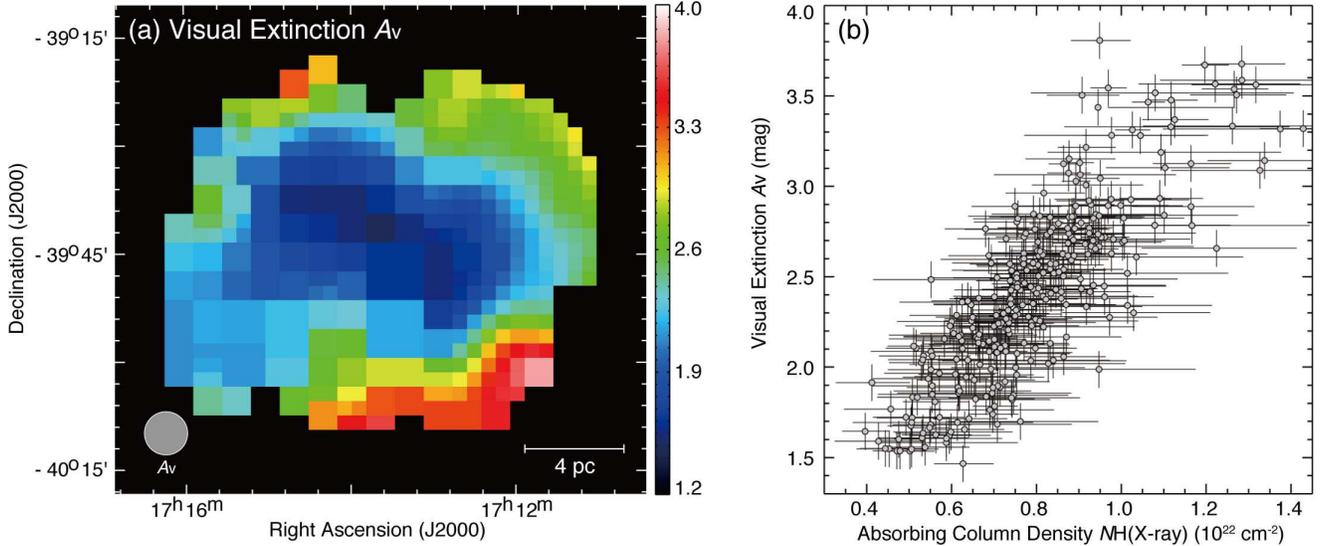}
\caption{(a) Distribution of the visual extinction $A_{\mathrm{v}}$ \citep{dobashi2005}. The grid separations are the same as Figure \ref{fig3}. (b) Correlation plot between the absorbing column density $N_{\mathrm{H}}$(X-ray) (in units of 10$^{22}$ cm$^{-2}$) and the visual extinction $A_{\mathrm{v}}$ (in units of mag). The error bars are given at a 90$\%$ confidence level in $N_{\mathrm{H}}$(X-ray).}
\label{fig4}
\end{center}
\end{figure*}%

We find that the global distribution of $N_{\rm{H}}$(X-ray) shows a shell-like structure in Figure \ref{fig3}a. In the southeast and the center, $N_{\rm{H}}$(X-ray) is as small as 0.4--0.5 $\times$10$^{22}$ cm$^{-2}$ (region in blue). On the other hand, the northeast and the northwest show medium values of $N_{\rm{H}}$(X-ray) = 0.7--1.0 $\times$ 10$^{22}$ cm$^{-2}$ (region in green), and the southwest and south show the largest $N_{\rm{H}}$(X-ray) with 1.1--1.4 $\times$10$^{22}$ cm$^{-2}$ (region in red). Therefore, the absorbing column density varies from 0.4$\times$10$^{22}$ cm$^{-2}$ to 1.4$\times$10$^{22}$ cm$^{-2}$ within the SNR. These values are mostly consistent with the previous studies with $XMM$-$Newton$ \citep[][typical angular resolution $\sim8\arcmin$]{cassamchenai2004}, while the angular resolution and the source coverage are better in the present study.

Figure \ref{fig4}a shows absorbing column density $N_{\rm{H}}$(X-ray) and visual extinction ($A_{\rm{V}}$ in unit of mag) derived from the Digitized Sky Survey I \citep[DSS;][]{dobashi2005}. The visual extinction is smoothed to the same binning with the X-rays. The values of $A_{\rm{V}}$ are roughly classified into the three levels; low values ($A_{\rm{V}}$ $<$ 2 mag in blue) in the center of the SNR, medium values (2 mag $<$ $A_{\rm{V}}$ $<$ 3 mag in green) in the northeast and northwest regions, and high values ($A_{\rm{V}}$ $>$ 3 mag in red) in the southwest and the south. The general trend of $A_{\rm{V}}$ is similar to that of $N_{\rm{H}}$(X-ray). Figure \ref{fig4}b shows a correlation plot between absorbing column density $N_{\rm{H}}$(X-ray) and visual extinction, showing a good correlation with a correlation coefficient of $\sim$0.83.

\begin{figure*}
\begin{center}
\includegraphics[width=180mm,clip]{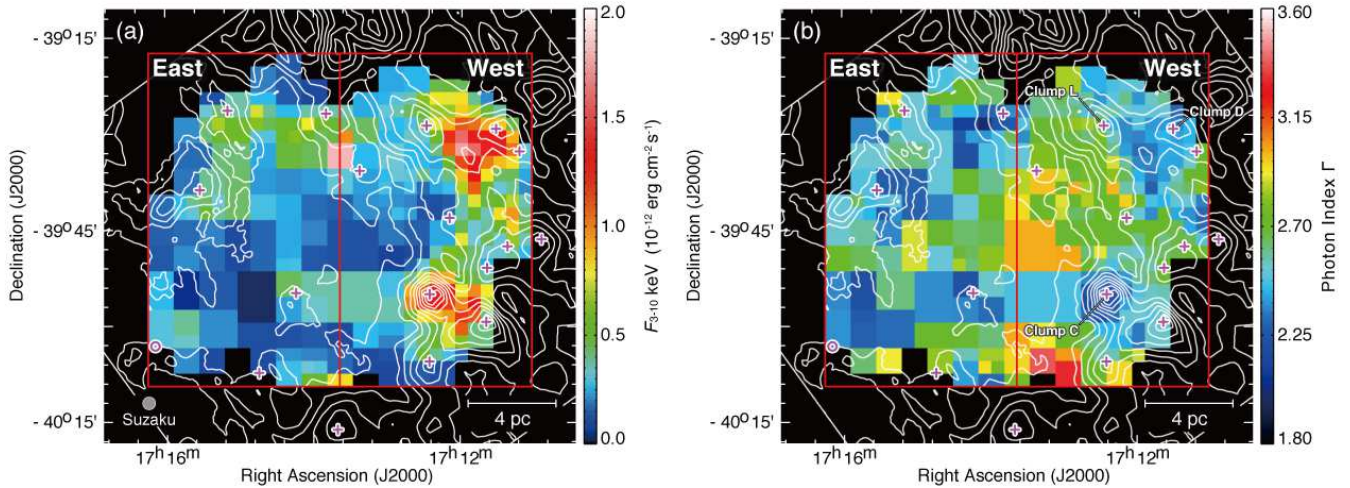}
\caption{Distribution of (a) the absorption-corrected flux $F_{3-10 \mathrm{keV}}$ and (b) photon index $\Gamma$ as Figure \ref{fig3}. The contours indicate the proton column density $N_{\mathrm{H}}$(H$_2$+H{\sc i}) in a velocity range from $-20$ to 2 km s$^{-1}$. The contour levels are 0.6, 0.75, 0.90, 1.1, 1.3, 1.5, 1.7, 1.9, and 2.1 $\times$ 10$^{22}$ cm$^{-2}$. The magenta crosses and circle are corresponded to the positions of CO and H{\sc i} clumps \citep{sano2013}.}
\label{fig5}
\end{center}
\end{figure*}%

\subsubsection{Photon index and Flux}
Figures \ref{fig3}b and \ref{fig3}c show the distributions of photon index $\Gamma$ and absorption-corrected flux $F_{3-10 \mathrm{keV}}$, respectively. Their relative errors and maximum errors are $\sim$6$\%$ and 13$\%$ in Figure \ref{fig3}b, and $\sim$7$\%$ and 23$\%$ in Figure \ref{fig3}c at the 90$\%$ confidence level, respectively. Like absorbing column density, these quantities show significant spatial variation. In particular, photon index is largest with $\Gamma$ $\sim$3 (in orange color) toward the center and the south. On the other hand, regions with a small photon index ($\Gamma < 2.4$) are distributed as islands inside the SNR from the north to the southeast and from the southwest to the northwest. Both regions have areas of $\sim430$ arcmin$^2$ and $\sim150$ arcmin$^2$, respectively. The absorption-corrected flux $F_{3-10 \mathrm{keV}}$ is similar to the 1--5 keV X-rays contours, especially at the brightest peaks in the northwest, west, and southwest. We also find another strong peak in the north. These flux excesses are not due to a systematic error and are possibly connected with the ISM distribution as discussed in Section \ref{comparisonISM}.

\begin{figure}
\begin{center}
\includegraphics[width=79mm,clip]{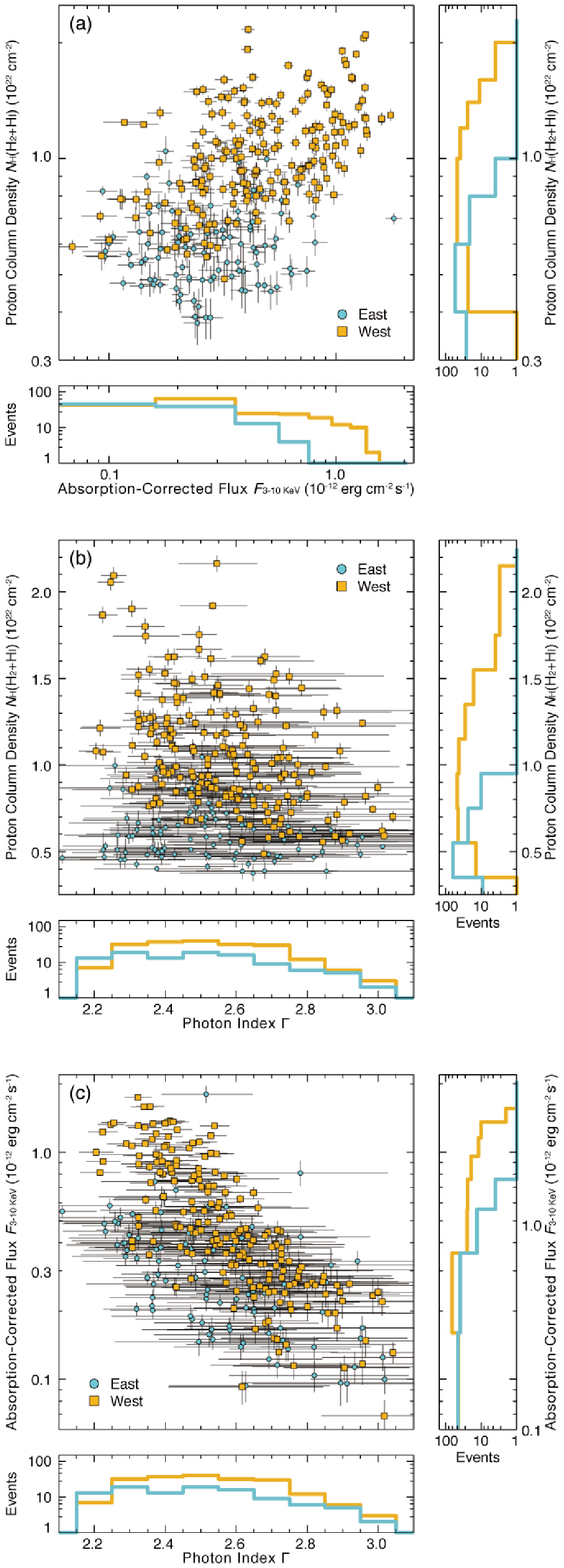}
\caption{Correlation plots for (a) the absorption-corrected flux $F_{3-10 \mathrm{keV}}$ vs. the proton column density $N_{\mathrm{H}}$(H$_2$+H{\sc i}), (b) the photon index $\Gamma$ vs. the proton column density $N_{\mathrm{H}}$(H$_2$+H{\sc i}), and (c) the photon index $\Gamma$ vs. the proton column density $N_{\mathrm{H}}$(H$_2$+H{\sc i}), respectively. The regions are defined as follow Figure \ref{fig5}: {\footnotesize${\bigcirc}$} for East, $\Box$ for West. The histograms are also shown at lower or right side of each plot.}
\label{fig6}
\end{center}
\end{figure}%

\subsection{Comparison with the ISM: The X-ray flux, photon index and the ISM}\label{comparisonISM}
Overlays of the interstellar gas at $V_{\rm{LSR}}$ = $-$20 to 2 km s$^{-1}$ with absorption-corrected flux $F_{3-10 \mathrm{keV}}$ and photon index $\Gamma$ are shown in Figure \ref{fig5}a and Figure \ref{fig5}b, respectively, where the ISM is total proton column density $N_{\rm{H}}$(H$_2$+H{\sc i}) including derived from the CO and H{\sc i}. The velocity range of the ISM associated is derived by \cite{moriguchi2005} and is supported by \cite{sano2013}. based on a good correlation between the ISM and X-rays enhanced by the interaction with the SNR. We also annotated the positions (center of gravity) of the molecular clumps (crosses) and the H{\sc i} clump (circle) defined by \cite{sano2013}. Additionally, we will hereinafter refer to the regions between the clumps as ``inter-clump\rq\rq{}. Figure \ref{fig5}a shows a trend that the X-rays are enhanced toward the regions with enhanced ISM in a pc scale. This trend is most significant in the west of the SNR, where the ISM is rich. On the other hand, in a sub-pc scale, we find that each bright spot of X-rays is lying around the CO and H{\sc i} clumps except for the molecular clump in southwest ($\alpha_{\mathrm{J2000}}$ = 17$^{\mathrm{h}}$12$^{\mathrm{m}}$25.3$^{\mathrm{s}}$, $\delta_{\mathrm{J2000}}$ = $-$39$^{\circ}$55$\arcmin$7.4$\arcsec$; named as ``clump C\rq{}\rq{}). These results are consistent with the previous study \citep{sano2013} and the present work has made it possible to evaluate the trend more quantitatively. Figure \ref{fig6}a shows a correlation plot between $F_{3-10 \mathrm{keV}}$ and the ISM density; the linear correlation coefficients (hereafter LCC) are $\sim$0.55 (for the whole), $\sim$0.19 (the eastern half, on the left side of $\alpha_{\mathrm{J2000}}$ $\sim$17$^{\mathrm{h}}$13$^{\mathrm{m}}$38$^{\mathrm{s}}$) and $\sim$0.54 (the western half, on the right side of $\alpha_{\mathrm{J2000}}$ $\sim$17$^{\mathrm{h}}$13$^{\mathrm{m}}$38$^{\mathrm{s}}$), respectively. The LCC$\sim$ 0.55 does not show strong correlation but the number of sample in the plot $\sim$300 indicates a positive correlation at a 5$\%$ confidence level according to a T-test, because the value of test statistic ($\sim$11.7) is greater than that of Student's t distribution ($\sim$1.97) \citep[e.g.,][]{taylor1982}.

In Figure \ref{fig5}b, we note that the regions having the hard spectrum are seen toward the enhanced ISM in the west, whereas the regions of the hard spectrum are also found toward the diffuse ISM in the eastern half. Most outstanding are the ISM peaks toward/around the two X-ray peaks in the northwest and in the southwest. In addition, we find four regions of the hard spectrum in the southeast where the ISM is diffuse. Figure \ref{fig6}b shows a correlation plot showing that the relationship between the ISM and photon index is different between the west and east. In the western half, the spectrum is hard when the ISM is dense and the spectrum is soft where the ISM is diffuse (LCC $\sim -0.41$). In the eastern half, the ISM is diffuse and the photon index shows variation (LCC $\sim -0.08$), where the ISM density is not apparently related to the variation of the photon index for the whole region LCC is estimated to be $\sim-0.16$. Additionally, we show correlation plots between the photon index $\Gamma$ and the absorption-corrected flux $F_{3-10 \mathrm{keV}}$ in Figure \ref{fig6}c. LCCs are $\sim -0.62$ (the whole), $\sim -0.53$ (the eastern half) and$\sim -0.81$ (the western half), respectively. The histograms toward the west and east regions show a similar trend to the photon index distribution with respected the ISM density, while the west region has much higher X-ray fluxes than the east. We shall discuss about these results later in Section \ref{efficient}.

We summarize the main aspects of the present analysis as follows (Figures \ref{fig5} and \ref{fig6});

\begin{enumerate}
\item In the western half of the SNR, the ISM density is high with significant H$_2$ and the X-rays are enhanced. In the eastern half of the SNR, the ISM density is low, being dominated by H{\sc i}, and the X-rays are weak. The overall correlation between the ISM density and the X-rays is however not significantly high with a correlation coefficient of $\sim$0.55 (Figure \ref{fig6}a).
\item The smallest X-ray photon index $< 2.4$ is seen around/toward the two CO peaks in the the western half, and toward the region of low ISM density in the eastern half (Figure \ref{fig5}b). The largest X-ray photon index around 3 is found toward the central region of the SNR as well as in the southern edge (Figure \ref{fig5}b).
\item The photon index shows a good correlation with the X-rays and the low photon index is seen toward regions of intense X-rays and vice versa (Figure \ref{fig6}c). There is an offset by 0.3 in this correlation between the eastern and western halves of the SNR in the sense that the X-rays are more intense in the western half, where the ISM is denser than in the eastern half.
\end{enumerate}

\section{Discussion}\label{section:discussion}

\subsection{Spatial variation of the absorbing column density}\label{absorbingcol}
The observations with $Suzaku$ XIS have enabled us to estimate detailed distributions of the absorbing column density $N_{\rm{H}}$(X-ray) and photon index $\Gamma$ in RX J1713.7$-$3946 at a low background level and high photon statics. Here in the discussion, we first discuss the absorbing column density $N_{\rm{H}}$(X-ray).

\begin{figure*}
\begin{center}
\includegraphics[width=180mm,clip]{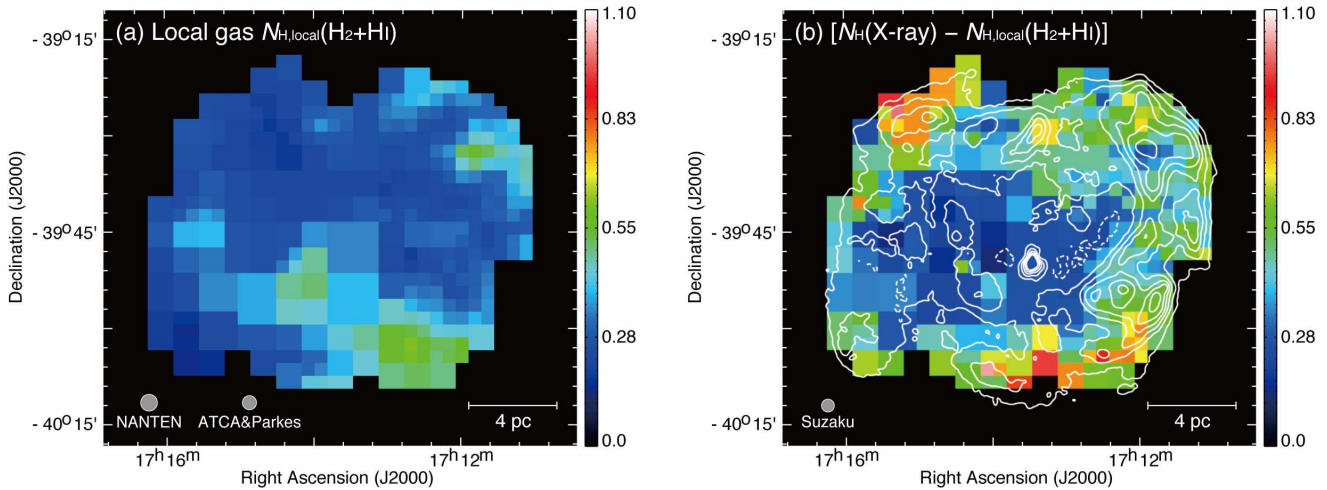}
\caption{(a) Distribution of the proton column density in local gas $N_{\mathrm{H, local}}$(H$_2$+H{\sc i}) estimated by using the CO and H{\sc i} data sets. (b) Distribution of the absorbing column density [$N_{\mathrm{H}}$(X-ray)$-$$N_{\mathrm{H, local}}$(H$_2$+H{\sc i})] overlaid with smoothed contours of the $Suzaku$ XIS mosaic images as shown in Figure \ref{fig3}. Both images are the same color scale.}
\label{fig7}
\end{center}
\end{figure*}

The spatial distribution of $N_{\rm{H}}$(X-ray) delineates the SNR shell as shown in Figure \ref{fig3}a. This distribution shows a good correspondence with visual extinction shown in Figure \ref{fig4}a. The regression in a straight line calculated from a least-squares fitting gives $N_{\rm{H}}$(X-ray) (cm$^{-2}$) = 3.0$\times$10$^{21}$$\cdot A_{\rm{V}}$ (magnitude) for the scatter plot in Figure \ref{fig4}b. The numerical factor is slightly larger than that of the conventional relation $N_{\rm{H}}$ (cm$^{-2}$) = 2.5$\times$10$^{21}$$\cdot A_{\rm{V}}$ (magnitude) \citep{jenkins1974}. This is understandable if we consider the distance of RX J1713.7$-$3946 is 1 kpc, since the visual extinction tends to be under-estimated for a distance larger than a few 100 pc by the foreground stars. It is not necessarily true that all $N_{\rm{H}}$(X-ray) is physically associated with the SNR. Here we need to take into account the contribution of the local gas between the SNR and the sun (Figure \ref{fig7}a). \cite{moriguchi2005} already estimated the foreground component $N_{\mathrm{H, local}}$(H$_2$+H{\sc i}) (Figure \ref{fig7}a). Figure \ref{fig7}b shows the distribution of absorbing column density [$N_{\mathrm{H}}$(X-ray)$-$$N_{\mathrm{H, local}}$(H$_2$+H{\sc i}) in the SNR, which gives a shell-like distribution of $N_{\rm{H}}$(X-ray) more clearly than Figure \ref{fig3}a. The absorption toward the center of the SNR is $\sim$0.2$\times$10$^{22}$ cm$^{-2}$, whereas that toward the outer boundary is shell-like with absorbing column density of $\sim$0.5--0.8$\times$10$^{22}$ cm$^{-2}$. This shell represents a cavity wall of the ISM created by the stellar wind of the SNR progenitor. The inside of the cavity is highly evacuated with density lower than $\sim1$ cm$^{-3}$ and the dense clumps in the cavity wall have higher density like $\sim$10$^2$--10$^4$ cm$^{-3}$ \citep[e.g.,][]{sano2010,inoue2012}.

We shall discuss the absorption toward the southeast-rim. In this region \cite{fukui2012} identified cold H{\sc i} gas that corresponds to the VHE $\gamma$-ray shell. The cold H{\sc i} with low spin temperature of $\sim$40 K has density around 100 cm$^{-3}$, less than $\sim$1000 cm$^{-3}$ threshold density for the collisional excitation of the CO emission. The proton column density of the cold H{\sc i} without CO emission is estimated to be $\sim$0.5$\times$10$^{22}$ cm$^{-2}$ from the H{\sc i} self-absorption in the southeast rim of RX J1713.7$-$3946 \citep{fukui2012}. The present absorption in the SNR calculated from X-rays has also column density of 0.4--0.5$\times$10$^{22}$ cm$^{-2}$ (Figure \ref{fig7}b) which is consistent with the cold H{\sc i}.

\begin{deluxetable*}{ccccccc}
\tabletypesize{\scriptsize}
\tablecaption{\textsc{Summary of the Correlation Coefficient in Scatter Plots}}
\tablewidth{0pt}
\tablehead{\multicolumn{2}{c}{Physical Parameters}&& \multicolumn{3}{c}{Correlation Coefficient (LCC)} & \\
\cline{1-2}\cline{4-6}
X-axis & Y-axis && Whole region & East half & West half & Related plot}
\startdata
$N_{\mathrm{H}}$(X-ray) & $A_V$ & & \phantom{0}0.83 &\nodata&\nodata& Figure \ref{fig5}b\\
$F_{3-10 \mathrm{keV}}$ & $N_{\mathrm{H}}$(H$_2$+H{\sc i}) & & \phantom{0}0.55 & \phantom{0}0.16 & \phantom{0}0.53 & Figure \ref{fig6}a\\
$\Gamma$ & $N_{\mathrm{H}}$(H$_2$+H{\sc i}) & & $-0.16$ & $-0.09$ & $-0.41$ & Figure \ref{fig6}b\\
$\Gamma$ & $F_{3-10 \mathrm{keV}}$ & & $-0.62$ & $-0.53$ & $-0.81$ & Figure \ref{fig6}c\\
$N_{\mathrm{H}}$(X-ray)$-$$N_{\mathrm{H, local}}$(H$_2$+H{\sc i}) & Blue-Shifted ISM & & \phantom{0}0.25 & \phantom{0}0.18 & \phantom{0}0.25 & Figure \ref{fig9}a\\
$N_{\mathrm{H}}$(X-ray)$-$$N_{\mathrm{H, local}}$(H$_2$+H{\sc i}) & Red-Shifted ISM & & \phantom{0}0.39 & \phantom{0}0.45 & \phantom{0}0.39 & Figure \ref{fig9}b
\enddata
\label{tab1}
\tablecomments{$N_{\mathrm{H}}$(X-ray): absorbing column density, $A_V$: visual extinction, $F_{3-10 \mathrm{keV}}$: absorption-corrected X-ray flux, $\Gamma$: photon index, $N_{\mathrm{H}}$(X-ray)$-$$N_{\mathrm{H, local}}$(H$_2$+H{\sc i}): proton column density.}
\end{deluxetable*}

\begin{figure*}
\begin{center}
\includegraphics[width=180mm,clip]{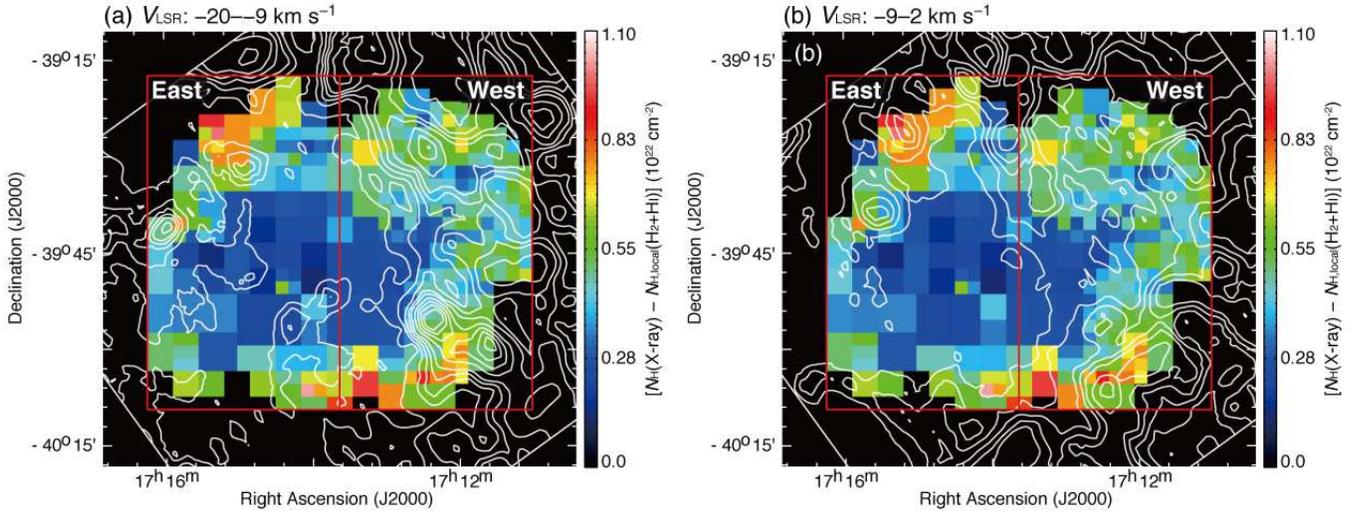}
\caption{Distribution of the absorbing column density [$N_{\mathrm{H}}$(X-ray)$-$$N_{\mathrm{H, local}}$(H$_2$+H{\sc i})] as Figure \ref{fig7} (c) superposed on the proton column density $N_{\mathrm{H}}$(H$_2$+H{\sc i}) in the velocity range from (a) $-20$ to $-9$ km s$^{-1}$ and from (b) $-9$ to 2 km s$^{-1}$, respectively. The contour levels are 0.4, 0.5, 0.6, 0.7, 0.9, 1.1, 1.3, and 1.5 $\times$ 10$^{22}$ cm$^{-2}$. The regions, East and West, where used for the correlation plots (see Figure \ref{fig9}).}
\label{fig8}
\end{center}
\end{figure*}

\begin{figure*}
\begin{center}
\includegraphics[width=180mm,clip]{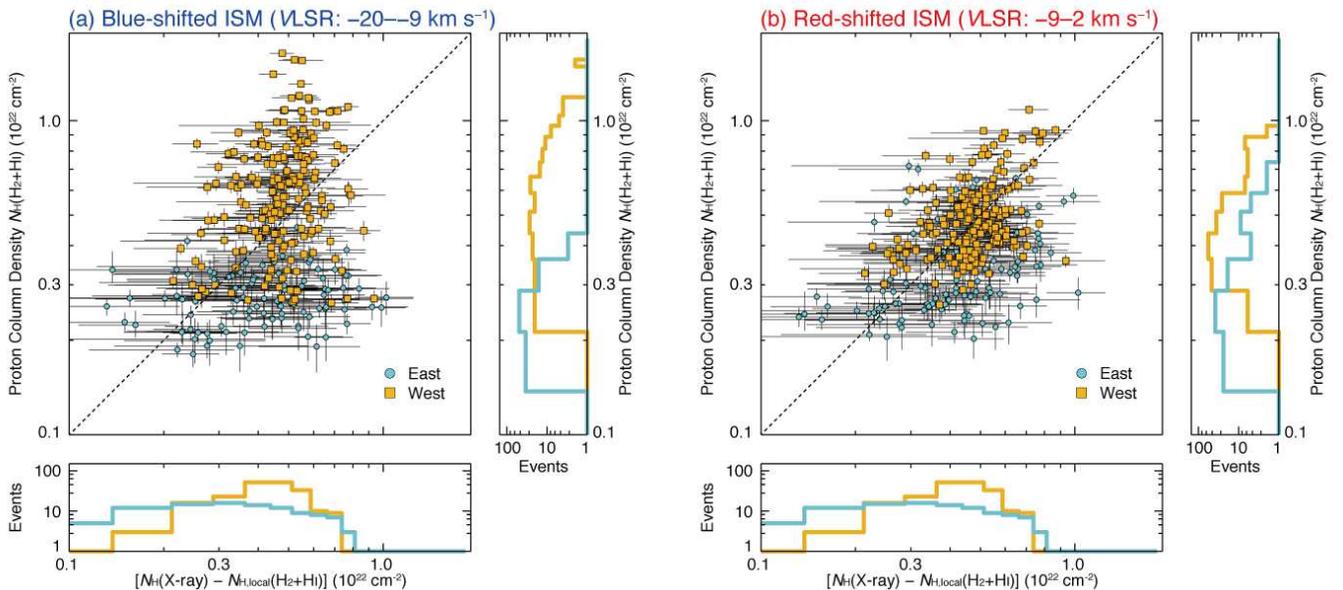}
\caption{Correlation plot between the absorbing column density [$N_{\mathrm{H}}$(X-ray)$-$$N_{\mathrm{H, local}}$(H$_2$+H{\sc i})] and the proton column density $N_{\mathrm{H}}$(H$_2$+H{\sc i}) in the velocity range from $-20$ to $-9$ km s$^{-1}$ in (a) and from $-9$ to 2 km s$^{-1}$ in (b), respectively. The dashed lines are the bisector for each panel. The histograms are also shown at lower or right side of each plot. All plots, the scale is square-root.}
\label{fig9}
\end{center}
\end{figure*}

We compared the absorbing column density [$N_{\rm{H}}$(X-ray) $-$ $N_{\rm{H\_local}}$(H$_2$+H{\sc i})] (Figure \ref{fig7}b) with $N_{\mathrm{H}}$(H$_2$+H{\sc i}), the proton column density physically associated with the X-ray emitting shell, in two velocity ranges of the interacting gas in Figures \ref{fig8} and \ref{fig9}. Figures \ref{fig8}a and \ref{fig8}b show [$N_{\rm{H}}$(X-ray) $-$ $N_{\rm{H\_local}}$(H$_2$+H{\sc i})] overlayed on the ISM proton column density $N_{\mathrm{H}}$(H$_2$+H{\sc i}) at $V_{\rm{LSR}}$ $\sim$ $-20$ to $-9$ km s$^{-1}$ and $\sim$ $-$9 to 2 km s$^{-1}$. This comparison shows that the ISM distribution has generally good correspondence with the X-ray absorbing column density. It is however to be noted that the blue-shifted ISM shows a poorer correlation than that of the red-shifted ISM with the X-ray absorbing column density as shown in Figure \ref{fig9}, scatter plots between the ISM and the X-ray absorption column density. In Figure \ref{fig9} we divide the plots into the red- and blue-shifted ISM. The blue-shifted ISM in the west shows a poor correlation with the X-ray absorbing column density of LCC$\sim$0.25, while the red-shifted ISM in the west has a higher correlation of LCC$\sim$0.39 than the blue-shifted ISM (see also Table \ref{tab1}). This indicates that the red-shifted ISM is located on the far-side of the SNR, and it is in the opposite sense to what is expected in an expanding motion of the shell-like ISM. The relative position of the ISM is explicable if the pre-existent cloud motion is dominant for the dense CO gas instead of expansion, as already suggested to explain the velocity distribution of the H{\sc i} self-absorption \cite{fukui2012}.

\subsection{Relationship between the X-ray flux, the photon index, and the X-ray absorption/the ISM}
\label{relationship}
\subsubsection{Shock-cloud interaction}
\cite{inoue2012} showed by magnetohydrodynamic (MHD) numerical simulations that the X-ray intensity is closely correlated with the ISM by the enhanced magnetic field around dense clumps due to turbulence in the shock-cloud interaction. The interaction creates correlation between the ISM and X-rays at a pc scale via the magnetic field. Observations by \cite{sano2010,sano2013} showed that pc-scale correlation is seen between the X-ray intensity and the clump mass interacting with the SNR blast waves, as well as anti-correlation between them in a sub-pc scale due to exclusion of the CR electrons in the dense clumps. The distributions of the X-ray intensity in Figure \ref{fig5}a at grid sizes of 2--8 arcmin (0.6--2.4 pc) show their interrelation at a pc scale; we see trend that the X-rays are enhanced in the west where the ISM is rich, and that the X-rays are depressed in the east where the ISM is poor. This is consistent with what is expected in the shock-cloud interaction scheme. The higher dispersion in Figure \ref{fig6}a may be in part ascribed to anti-correlation at a sub-pc scale, consistent with the suggestion by \cite{sano2010,sano2013}.

\subsubsection{Efficient cosmic-ray acceleration}\label{efficient}
Based on the present results on the photon index (Figure \ref{fig3}b), we discuss the efficient cosmic-ray acceleration in RX J1713.7$-$3946. As shown by the previous works \citep{takahashi2008,tanaka2008}, the photon index in the 1--10keV range reflects rolloff energy $\varepsilon_0$ of the synchrotron X-rays. The small (large) photon index corresponds to large (small) rolloff energy. According to the standard DSA scheme, the rolloff energy of synchrotron photons $\varepsilon_0$ is given as follows when the synchrotron cooling is effective \citep{zirakashvili2007},
\begin{eqnarray}
\varepsilon_0 = 0.55 \times (v_\mathrm{sh}\:/\:3000 \;\mathrm{km\; s^{-1}})^2 \:\eta^{-1} \;\mathrm{(keV)},
\label{eq01}
\end{eqnarray}
where $v_\mathrm{sh}$ is the shock speed and $\eta = B^2/\delta B^2$ ($>$1) a gyro-factor, the degree of magnetic field fluctuations. In particular the limit of $\eta$ = 1 is called as Bohm limit corresponding to highly turbulent conditions and, accordingly, the rolloff energy is likely determined by the shock speed and turbulence.

The present results suggest that in the western half of the SNR, where the ISM is rich, the shock-cloud interaction is effective and that turbulence is enhanced around dense ISM clumps. On the other hand, in the eastern half of the SNR where the ISM is poor the DSA alone is mainly working without shock-cloud interaction. \cite{sano2013} presented that the innermost cavity of 3--4 pc radius is surrounded by the ISM shell consisting of the dense clumps and the inter-clump gas. The dense clumps have density of 10$^2$ to 10$^4$ cm$^{-3}$ \citep{sano2010,sano2013} and are mainly distributed in the western half toward the Galactic plane. The inter-clump gas has density $<2$ cm$^{-3}$, as given by the upper limit from no thermal X-rays \citep{takahashi2008}, and we shall adopt density of the inter-clump gas to be 1 cm$^{-3}$ for discussion. We assume that density in the cavity is $\sim$0.25 cm$^{-3}$, the same with that estimated toward Gum nebula where the gas is swept up by the strong stellar winds from $\zeta$ Pup \citep{wallerstein1971,gorenstein1974}. In this scheme, DSA is efficiently taking place in the cavity and the accelerated CRs are injected into the ISM shell.

In the western half of the SNR outside the center (Figure \ref{fig3}d), where the shock-cloud interaction is working \citep[e.g.,][]{sano2010,inoue2012,sano2013}, the magnetic field may be amplified up to $\sim$1 mG as indicated by the short time variation of $\sim$1 yr in the synchrotron X-rays by $Chandra$, where the Bohm limit ($\eta$ = 1) is a good approximation \citep{uchiyama2007}. On the other hand, in the eastern half the shock speed is not much decelerated in the lower density. Because the shock speed $v_\mathrm{sh}$ is proportional to 1/$\sqrt{n}$ and decreases in the dense ISM, where $n$ is the average number density of the ISM. The average ISM density in the eastern half is about 1/4 of that in the western half from the CO/H{\sc i} observations and the shock speed $v_\mathrm{sh}$ becomes larger by a factor of 2. Then, $\eta$ in the southeast becomes larger by a factor of 4, but the term (the shock speed $v_\mathrm{sh}$)$^2$ is more effective. The trend in the photon index is thus largely explained in the DSA scheme by considering the effect of the ISM gas, and in the both regions the rolloff energy becomes larger and the photon index smaller as $\Gamma < 2.4$.

In Figure \ref{fig5}b, the center of the SNR (Figure \ref{fig3}b) having the large photon index $\Gamma > 2.8$ is shown by orange color. We suggest that the electrons there have lower rolloff energy due to synchrotron cooling over the last 1000 yrs; for magnetic field of 10 $\mu$G the CR electron cooling time at 10 keV is small as $\sim$500 yrs, whereas that at 1 keV is large as $\sim$1500 yrs, leading to low rolloff energy in the central 3--4 pc. We also note that such spectral softening can be ascribed to the energy loss due to adiabatic expansion \citep[e.g.,][]{kishishita2013}.

One would expect that the photon index becomes even larger toward the denser regions, if only $v_\mathrm{sh}$ determines the photon index. Interestingly, the photon index becomes small toward the clump C and the region between the two dense clumps, D and L in Figure \ref{fig5}b. It is also notable that the X-rays are enhanced toward these regions having small photon indexes. In the shock-cloud interaction scheme, the magnetic field is amplified in the interacting region, leading to higher synchrotron loss and smaller rolloff energy. This can lead to lower X-ray intensity and lower rolloff energy toward the dense clumps, while obviously we need a more elaborate work to quantitatively affirm this. It remains to be explored if the DSA scheme can accommodate the high X-ray intensity, the high ISM density, and the small photon index consistently. Also we here suggest that some additional acceleration mechanism may possibly be working to accelerate CR particles in the west where the ISM is rich. Such mechanisms possibly include the 2nd order Fermi acceleration \citep{fermi1949}, magnetic reconnection in the turbulent medium \citep{hoshino2012}, reverse shock acceleration \citep[e.g.,][]{ellison2001}, non-liner effect of DSA \citep[e.g.,][]{malkov2001} and so on.

To summarize, we suggest that the ISM and its distribution have a significant impact on the scheme of the particle acceleration via the shock-cloud interaction. This suggests the tight relationship between the SNR and the ISM. The present study explored the CR electron behavior into significant details not achieved in the previous works, and presented that the ISM density affects significantly the CR electron rolloff energy. In particular, dense molecular clumps excites turbulent in the shock waves, amplifying the magnetic field. The CR electron rolloff energy is increased in spite of such amplified field, and we suggest that additional electron acceleration may be taking place in such turbulent regimes around the dense clumps.

In near future, Cherenkov Telescope Array (CTA) will provide information on the spectral distribution of $\gamma$-rays over the SNR at 1 arcmin resolution and will allow us to investigate the spectra and acceleration of both the CR protons and electrons. In addition, the soft X-ray spectrometer (SXS) of ASTRO-H will probe yet undetected thermal X-rays by sensitive high-energy resolution spectroscopy in RX J1713.7$-$3946. We can then probe the fraction of the energy of the SNR blast waves used for cosmic-ray acceleration. And the hard X-ray imager (HXI) of ASTRO-H will resolve photon index distribution at 10 keV or higher, possibly allowing us to constrain electron energy spectrum models by comparing that below 10 keV \citep[e.g.,][]{yamazaki2014}. These future instruments will enable more accurate measurements of the efficient cosmic-ray acceleration.

\section{Conclusions}\label{section:Conclusion}

We summarize the present work as follows.

\begin{enumerate}
\item We have estimated the spatial distribution of absorbing column density, photon index, and absorption-corrected flux (3--10 keV) comparable to the scale of the ISM distribution, a few arcmin, by using $Suzaku$ archival data with low background.
\item The X-ray flux shows enhancement toward the dense ISM. This is consistent with the shock-cloud interaction model by \citep{inoue2012} that X-rays become bright around the dense cloud cores due to turbulence amplification of magnetic field.
\item Photon index shows a large variation within the SNR from $\Gamma$ = 2.1--2.9. The photon index shows smallest values around the dense regions of cloud cores as well as toward diffuse regions with no molecular gas. This trend can be described as a rolloff energy variation of CR electrons \citep{zirakashvili2007}. We present possible parameters to explain the variation of the rolloff energy under the DSA scheme. The enhanced intensity and harder spectra of the X-rays toward the dense clumps may require additional electron acceleration toward the dense clumps possibly via magnetic reconnection or other mechanism incorporating magnetic turbulence.
\item The absorbing column density shows a good correlation with the visual extinction. We found that the southeast-rim identified by H{\sc i} self-absorption, which shows enhanced VHE $\gamma$-rays has a clear counterpart in the X-ray absorption too, lending new support to the H{\sc i} self-absorption interpretation.
\end{enumerate}

\acknowledgments
NANTEN2 is an international collaboration of 10 universities; Nagoya University, Osaka Prefecture University, University of Cologne, University of Bonn, Seoul National University, University of Chile, University of New South Wales, Macquarie University, University of Sydney, and University of ETH Zurich. This work was financially supported by a grant-in-aid for Scientific Research (KAKENHI, No. 24$\cdot$10082, No. 21253003, No. 23403001, No. 22540250, No. 22244014, No. 23740149, No. 23740154 (T.I.), No. 22740119, and No. 24224005) from the Ministry of Education, Culture, Sports, Science and Technology of Japan (MEXT). This work was also financially supported by the Young Research Overseas Visits Program for Vitalizing Brain Circulation (R2211) and the Institutional Program for Young Researcher Overseas Visits (R29) by Japan Society for the Promotion of Science (JSPS) and by the Mitsubishi Foundation and by the grant-in-aid for Nagoya University Global COE Program, ``Quest for Fundamental Principles in the Universe: From Particles to the Solar System and the Cosmos\rq{}\rq{}, from MEXT. This research made use of data obtained from Data ARchives and Transmission System (DARTS), provided by Center for Science-satellite Operation, and Data Archive (C-SODA) at ISAS/JAXA.


\begin{thebibliography}{99}
\bibitem[Abdo et al.(2011)]{abdo2011} Abdo, A.~A., Ackermann, M., Ajello, M., et al.\ 2011, \apj, 734, 28 
\bibitem[Acero et al.(2009)]{acero2009} Acero, F., Ballet, J., Decourchelle, A., et al.\ 2009, \aap, 505, 157 
\bibitem[Aharonian et al.(2004)]{aharonian2004} Aharonian, F.~A., Akhperjanian, A.~G., Aye, K.-M., et al.\ 2004, \nat, 432, 75 
\bibitem[Aharonian et al.(2005)]{aharonian2005} Aharonian, F., Akhperjanian, A.~G., Bazer-Bachi, A.~R., et al.\ 2005, \aap, 437, L7
\bibitem[Aharonian et al.(2006a)]{aharonian2006a} Aharonian, F., Akhperjanian, A.~G., Bazer-Bachi, A.~R., et al.\ 2006, \apj, 636, 777 
\bibitem[Aharonian et al.(2006b)]{aharonian2006b} Aharonian, F., Akhperjanian, A.~G., Bazer-Bachi, A.~R., et al.\ 2006, \aap, 449, 223 
\bibitem[Aharonian et al.(2007)]{aharonian2007a} Aharonian, F., Akhperjanian, A.~G., Bazer-Bachi, A.~R., et al.\ 2007, \aap, 464, 235 
\bibitem[Aharonian et al.(2007)]{aharonian2007b} Aharonian, F., Akhperjanian, A.~G., Bazer-Bachi, A.~R., et al.\ 2007, \apj, 661, 236
\bibitem[Bertsch et al.(1993)]{bertsch1993} Bertsch, D.~L., Dame, T.~M., Fichtel, C.~E., et al.\ 1993, \apj, 416, 587 
\bibitem[Berezhko \& V{\"o}lk(2006)]{berezhko2006} Berezhko, E.~G., \& V{\"o}lk, H.~J.\ 2006, \aap, 451, 981 
\bibitem[Berezhko \& V{\"o}lk(2008)]{berezhko2008} Berezhko, E.~G., \& V{\"o}lk, H.~J.\ 2008, \aap, 492, 695 
\bibitem[Berezhko \& V{\"o}lk(2010)]{berezhko2010} Berezhko, E.~G., \& V{\"o}lk, H.~J.\ 2010, \aap, 511, A34 
\bibitem[Butt et al.(2001)]{butt2001} Butt, Y.~M., Torres, D.~F., Combi, J.~A., Dame, T., \& Romero, G.~E.\ 2001, \apjl, 562, L167
\bibitem[Cassam-Chena{\"i} et al.(2004)]{cassamchenai2004} Cassam-Chena{\"i}, G., Decourchelle, A., Ballet, J., et al.\ 2004, \aap, 427, 199 
\bibitem[Dickey \& Lockman(1990)]{dickey1990} Dickey, J.~M., \& Lockman, F.~J.\ 1990, \araa, 28, 215
\bibitem[Dobashi et al.(2005)]{dobashi2005} Dobashi, K., Uehara, H., Kandori, R., et al. 2005, \pasj, 57, 1
\bibitem[Ellison(2001)]{ellison2001} Ellison, D.~C.\ 2001, \ssr, 99, 305 
\bibitem[Ellison \& Vladimirov(2008)]{ellison2008} Ellison, D.~C., \& Vladimirov, A.\ 2008, \apjl, 673, L47 
\bibitem[Ellison et al.(2010)]{ellison2010} Ellison, D.~C., Patnaude, D.~J., Slane, P., \& Raymond, J.\ 2010, \apj, 712, 287 
\bibitem[Ellison et al.(2012)]{ellison2012} Ellison, D.~C., Slane, P., Patnaude, D.~J., \& Bykov, A.~M.\ 2012, \apj, 744, 39 
\bibitem[Enomoto et al.(2002)]{enomoto2002} Enomoto, R., Tanimori, T., Naito, T., et al.\ 2002, \nat, 416, 823 
\bibitem[Fang et al.(2009)]{fang2009} Fang, J., Zhang, L., Zhang, J.~F., Tang, Y.~Y., \& Yu, H.\ 2009, \mnras, 392, 925 
\bibitem[Fermi(1949)]{fermi1949} Fermi, E.\ 1949, Physical Review, 75, 1169
\bibitem[Fukui et al.(2003)]{fukui2003} Fukui, Y., Moriguchi, Y., Tamura, K., et al.\ 2003, \pasj, 55, L61 
\bibitem[Fukui et al.(2012)]{fukui2012} Fukui, Y., Sano, H., Sato, J., et al.\ 2012, \apj, 746, 82 
\bibitem[Fukui(2013)]{fukui2013} Fukui, Y.\ 2013, in Astrophysics and Space Science Proc. 34, 2nd Session of the Sant Cugat Forum on Astrophysics, ed. Diego F. Torres $\&$ O. Reimer (Berlin: Springer), 249
\bibitem[Gabici \& Aharonian(2014)]{gabici2014} Gabici, S., \& Aharonian, F.~A.\ 2014, \mnras, 445, L70
\bibitem[Giacalone \& Jokipii(2007)]{giacalone2007} Giacalone, J., \& Jokipii, J.~R.\ 2007, \apjl, 663, L41
\bibitem[Gorenstein et al.(1974)]{gorenstein1974} Gorenstein, P., Harnden, F.~R., Jr., \& Tucker, W.~H.\ 1974, \apj, 192, 661
\bibitem[Hiraga et al.(2005)]{hiraga2005} Hiraga, J.~S., Uchiyama, Y., Takahashi, T., \& Aharonian, F.~A.\ 2005, \aap, 431, 953
\bibitem[Huang et al.(2007)]{huang2007} Huang, C.-Y., Park, S.-E., Pohl, M., \& Daniels, C.~D.\ 2007, Astroparticle Physics, 27, 429 
\bibitem[Hoshino(2012)]{hoshino2012} Hoshino, M.\ 2012, Physical Review Letters, 108, 135003
\bibitem[H.E.S.S.~Collaboration et al.(2011)]{hess2011} H.E.S.S.~Collaboration, Abramowski, A., Acero, F., et al.\ 2011, \aap, 531, A81
\bibitem[Inoue et al.(2009)]{inoue2009} Inoue, T., Yamazaki, R., \& Inutsuka, S.-i.\ 2009, \apj, 695, 825
\bibitem[Inoue et al.(2012)]{inoue2012} Inoue, T., Yamazaki, R., Inutsuka, S.-i., \& Fukui, Y.\ 2012, \apj, 744, 71 
\bibitem[Ishisaki et al.(2007)]{ishisaki2007} Ishisaki, Y., Maeda, Y., Fujimoto, R., et al.\ 2007, \pasj, 59, 113 
\bibitem[Jenkins \& Savage(1974)]{jenkins1974} Jenkins, E.~B., \& Savage, B.~D.\ 1974, \apj, 187, 243 
\bibitem[Taylor(1982)]{taylor1982} Taylor J. R. 1982 An introduction to error analysis (2nd ed.; California: USB)
\bibitem[Kishishita et al.(2013)]{kishishita2013} Kishishita, T., Hiraga, J., \& Uchiyama, Y.\ 2013, \aap, 551, A132
\bibitem[Koyama et al.(1997)]{koyama1997} Koyama, K., Kinugasa, K., Matsuzaki, K., et al.\ 1997, \pasj, 49, L7 
\bibitem[Li et al.(2011)]{li2011} Li, H., Liu, S., \& Chen, Y.\ 2011, \apjl, 742, L10 
\bibitem[Lucek \& Bell(2000)]{lucek2000} Lucek, S.~G., \& Bell, A.~R.\ 2000, \mnras, 314, 65
\bibitem[Malkov \& O'C Drury(2001)]{malkov2001} Malkov, M.~A., \& O'C Drury, L.\ 2001, Reports on Progress in Physics, 64, 429
\bibitem[Malkov et al.(2005)]{malkov2005} Malkov, M.~A., Diamond, P.~H., \& Sagdeev, R.~Z.\ 2005, \apjl, 624, L37 
\bibitem[Maxted et al.(2012)]{maxted2012} Maxted, N.~I., Rowell, G.~P., Dawson, B.~R., et al.\ 2012, \mnras, 422, 2230 
\bibitem[McClure-Griffiths et al.(2005)]{mccluregriffiths2005} McClure-Griffiths, N. M., Dickey, J. M., Gaensler, B. M., Green, A. J., Haverkorn, M., \& Strasser, S. 2005, \apjs, 158, 178.
\bibitem[Moriguchi et al.(2005)]{moriguchi2005} Moriguchi, Y., Tamura, K., Tawara, Y., et al.\ 2005, \apj, 631, 947 
\bibitem[Moraitis \& Mastichiadis(2007)]{moraitis2007} Moraitis, K., \& Mastichiadis, A.\ 2007, \aap, 462, 173
\bibitem[Morlino et al.(2009)]{morlino2009} Morlino, G., Amato, E., \& Blasi, P.\ 2009, \mnras, 392, 240 
\bibitem[Muraishi et al.(2000)]{muraishi2000} Muraishi, H., Tanimori, T., Yanagita, S., et al.\ 2000, \aap, 354, L57
\bibitem[Pannuti et al.(2003)]{pannuti2003} Pannuti, T.~G., Allen, G.~E., Houck, J.~C., \& Sturner, S.~J.\ 2003, \apj, 593, 377 
\bibitem[Pfeffermann \& Aschenbach(1996)]{pfeffermann1996} Pfeffermann, E., \& Aschenbach, B. 1996, in Roentgenstrahlung from the Universe, MPE Report 263, ed. H. U. Zimmermann, J. H. Tr\"{u}mper, \& H. Yorke, 267
\bibitem[Reimer \& Pohl(2002)]{reimer2002} Reimer, O., \& Pohl, M.\ 2002, \aap, 390, L43 
\bibitem[Sano et al.(2010)]{sano2010} Sano, H., Sato, J., Horachi, H., et al.\ 2010, \apj, 724, 59 
\bibitem[Sano et al.(2013)]{sano2013} Sano, H., Tanaka, T., Torii, K., et al.\ 2013, \apj, 778, 59 
\bibitem[Slane et al.(1999)]{slane1999} Slane, P., Gaensler, B.~M., Dame, T.~M., et al.\ 1999, \apj, 525, 357 
\bibitem[Takahashi et al.(2008)]{takahashi2008} Takahashi, T., Tanaka, T., Uchiyama, Y., et al.\ 2008, \pasj, 60, 131 
\bibitem[Tanaka et al.(2008)]{tanaka2008} Tanaka, T., Uchiyama, Y., Aharonian, F.~A., et al.\ 2008, \apj, 685, 988 
\bibitem[Uchiyama et al.(2003)]{uchiyama2003} Uchiyama, Y., Aharonian, F.~A., \& Takahashi, T.\ 2003, \aap, 400, 567 
\bibitem[Uchiyama et al.(2005)]{uchiyama2005} Uchiyama, Y., Aharonian, F.~A., Takahashi, T., et al.\ 2005, High Energy Gamma-Ray Astronomy, 745, 305
\bibitem[Uchiyama et al.(2007)]{uchiyama2007} Uchiyama, Y., Aharonian, F.~A., Tanaka, T., Takahashi, T., \& Maeda, Y.\ 2007, \nat, 449, 576
\bibitem[Wallerstein \& Silk(1971)]{wallerstein1971} Wallerstein, G., \& Silk, J.\ 1971, \apj, 170, 289  
\bibitem[Wang et al.(1997)]{wang1997} Wang, Z.~R., Qu, Q.-Y., \& Chen, Y.\ 1997, \aap, 318, L59 
\bibitem[Weaver et al.(1977)]{weaver1977} Weaver, R., McCray, R., Castor, J., Shapiro, P., \& Moore, R.\ 1977, \apj, 218, 377 
\bibitem[Yamazaki et al.(2014)]{yamazaki2014} Yamazaki, R., Ohira, Y., Sawada, M., \& Bamba, A.\ 2014, Research in Astronomy and Astrophysics, 14, 165
\bibitem[Yuan et al.(2011)]{yuan2011} Yuan, Q., Liu, S., Fan, Z., Bi, X., \& Fryer, C.~L.\ 2011, \apj, 735, 120 
\bibitem[Zhang \& Yang(2011)]{zhang2011} Zhang, L., \& Yang, C.~Y.\ 2011, \pasj, 63, 89 
\bibitem[Zirakashvili \& Aharonian(2007)]{zirakashvili2007} Zirakashvili, V.~N., \& Aharonian, F.\ 2007, \aap, 465, 695 
\bibitem[Zirakashvili \& Aharonian(2010)]{zirakashvili2010} Zirakashvili, V.~N., \& Aharonian, F.~A.\ 2010, \apj, 708, 965 
\end{thebibliography}
\end{document}